\documentclass[11pt]{article}
\usepackage{amsmath,amsfonts,amssymb,amsthm,amstext,amscd,eucal,srcltx}
\usepackage{amssymb}
\usepackage{latexsym}

\def \beq  {\begin{equation}}
\def \eeq  {\end{equation}}
\def \beqar {\begin{eqnarray}}
\def \eeqar {\end{eqnarray}}
\def\sqr#1#2{{\vcenter{\vbox{\hrule height.#2pt
\hbox{\vrule width.#2pt height#1pt \kern#1pt
\vrule width.#2pt}\hrule height.#2pt}}}}

\def\taubis{\mathcal{T}}

\begin{document}
%%%%%%%%%%%%%%%%%%%%%%%%%%%%%%%%%%%%%%%%%%%%%%%
\def \CMP {{Commun. Math. Phys.}}
\def \PRL {{Phys. Rev. Lett.}}
\def \PL {{Phys. Lett.}}
\def \NPBProc {{Nucl. Phys. B (Proc. Suppl.)}}
\def \NP {{Nucl. Phys.}}
\def \RMP {{Rev. Mod. Phys.}}
\def \JGP {{J. Geom. Phys.}}
\def \CQG {{Class. Quant. Grav.}}
\def \MPL {{Mod. Phys. Lett.}}
\def \IJMP {{ Int. J. Mod. Phys.}}
\def \JHEP {{JHEP}}
\def \PR {{Phys. Rev.}}
\def \JMP {{J. Math. Phys.}}
\def \GRG{{Gen. Rel. Grav.}}
%%%%%%%%%%%%%%%%%%%%%%%%%%%%%%%%%%%%%%%%%%%%%%%
%\renewcommand{\theequation}{\thesection.\arabic{equation}}
\fontfamily{cmr}\fontsize{11pt}{17.2pt}\selectfont
%%%%%%%%%%%%%%%%%%%%%%%%%%%%%%%%%%%%%%%%%%%%%%%
\begin{titlepage}
\null\vspace{-62pt} \pagestyle{empty}
\begin{center}
%\rightline{CCNY-HEP-08/5}
%\rightline{November 2008}
\vspace{1truein}
% {\Large\bfseries
% {Title}}\\
\vskip .1in
 {\Large\bfseries
 {Non linear Fierz-Pauli theory from torsion and bigravity}}

%{\Large\bfseries OR: Towards a consistent infrared modification of gravity}\\
%~\\
%{\Large\bfseries String Theories}\\
%\vskip .1in
%{\Large\bfseries ~}\\
%%%%%%%%%%%%%%%%%%%%%%%%%%%%%%%%%%%%%%%%%%%%%%%%%

\vspace{.5in} {\large C. DEFFAYET$^{~a,}$\footnote{E-mail: deffayet@iap.fr}, S. RANDJBAR-DAEMI$^{~b,}$\footnote{E-mail: seif@ictp.trieste.it}}\\
\vspace{.1in}{\itshape $^a$ APC, UMR 7164 (CNRS, Universit\'e Paris 7, CEA, Observatoire de Paris),\\ 10 rue Alice Domon et L\'eonie Duquet,
 75205 Paris Cedex 13, France.}\\
\vspace {.1in}{\itshape $^b$The Abdus Salam International Centre
for Theoretical Physics, Trieste, Italy}\\

\vspace{.4in}
%%%%%%%%%%%%%%%%%%%%%%%%%%%%%%%%%%%%%%%%%%%%%%%%%%%%%%%%%%%%
\centerline{\large\bf
}
\end{center}
 The non linear  aspects of a recently proposed model of massive spin-2 particles with propagating torsion are studied. 
 We obtain a nonlinear equation which reduces at linear order to a generalized Fierz-Pauli equation in any background space-time with or without a vanishing torsion. We contrast  those results with properties of a class of bigravity theories in an arbitrary background  Einstein manifold. It is known that the non perturbative spectrum of the bigravity model has 8 propagating physical degrees of freedom. This is identical to the physical propagating degrees of freedom of the massive spin-2 torsion model at the linearized order. The obtained non linear version of the Fierz-Pauli field equations, however,  contains terms absent in the bigravity case which indicates that the curved space generalization of the unique flat space space Fierz-Pauli equation is not unique.   Moreover, in the torsion massive gravity model the Fierz-Pauli field appears as a derivative of fundamental fields. This, however, does not generate any unwanted pole once coupled to some external sources.

\end{titlepage}
%%%%%%%%%%%%%%%%%%%%%%%%%%%%%%%%%%%%%%%%%%%%%%%%%%%%%%
\pagestyle{plain} \setcounter{page}{2}
\section{Introduction}

Infrared modifications of gravity may be a possibility to address some of the outstanding issues in cosmology.  A line of thought, that paved the way to many recent developpments along  this idea, started with the DGP model and its interesting cosmology \cite{Dvali:2000hr,DGPcosmo}. In the DGP model, large distance modifications of gravity arises because the usual graviton is a resonance of massive modes,  and many subsequent developments concentrated on various properties of "massive gravity" that are also found in the DGP model. In fact, the simplest way to modify gravity at large distance, would be to give a mass to the graviton. The consistent theory of a non self-interacting massive graviton is well know to be the 
Fierz-Pauli theory \cite{Fierz:1939ix} (henceforth FP theory). Extensive studies of this model and its non linear completion have exhibited many interesting features.
 First, it is known that Fierz-Pauli theory suffers from the van Dam-Veltman-Zakharov discontinuity \cite{vanDam:1970vg,Zakharov:1970cc}, stating, roughly speaking, that the massive theory does not behave as linearized General Relativity when the mass of the graviton is sent continuously to zero. This picture was shown to change drastically in the non linear theory, where, e.g., a critical distance scale, known as the Vainshtein radius \cite{Vainshtein:1972sx}, was found to appear around a massive body, below which gravity behaves effectively as in the massless case; while for distances above the Vainshtein radius, gravity is modified {\it \`a la} Yukawa as in the non interacting Fierz-Pauli theory \cite{Vainshtein:1972sx,Babichev:2009us}. This  "Vainshtein mechanim" is also relevant for 
other models such as the DGP model \cite{Deffayet:2001uk}, but also simpler scalar-tensor models \cite{ST}, some of which can be obtained 
from more complicated constructions having also to do with massive gravity \cite{Creminelli:2005qk,Deffayet:2005ys,DL}.
Despite such interesting  aspects, the  Fierz-Pauli model is limited by consistency problems,  such as the presence of a sixth spin zero  propagating ghost mode known as Boulware Deser mode \cite{Boulware:1973my}. This mode is not present at the level of linear equations of motion (or quadratic action), but is generically seen to propagate as soon as one non-linearly completes the theory and sticks with a Lorentz invariant theory (see, however, \cite{DRG} for a recent proposal aiming at circomventing this problem).

Recently, a new approach to massive gravity has been proposed \cite{Nair:2008yh,Nikiforova:2009qr} using a class of models in which the connection and vierbein are treated independently \cite{Hayashi:1979wj,Hayashi:1980ir,Hayashi:1980qp,torsion}. An interesting feature of these models is that unlike the FP theory and its various extensions the massive graviton originates from the torsion rather than from a second rank symmetric metric tensor.  In this work, we would like to underline that the models studied in Refs. \cite{Nair:2008yh,Nikiforova:2009qr} provide a new and largely unexplored way to build a non linear theory of massive gravity. With this aim we compare properties (some of which unnoticed before) of two models, one which is the model of Refs. \cite{Nair:2008yh,Nikiforova:2009qr} and that will be called here torsion massive gravity, and another model which is a certain class of bigravity theory \cite{Isham:gm,Salam:1976as,Isham:1977rj,Damour:2002ws}. Both models will be seen to have one massive and one massless spin-2 propagating graviton\footnote{ Linear fluctuations around any Einstein background in the  torsion massive gravity include also a massive spin zero mode while the bigravity model includes an extra spin-zero mode at a non perturbative level.}, as well as to be free from ghost modes in the background of any curved Einstein manifold, at least, for what concerns the new model of Refs. \cite{Nair:2008yh,Nikiforova:2009qr}, for weak enough curvature and at quadratic level of the action. However, the latter model will be seen to differ from the former. Indeed, in the torsion massive gravity, the field equations for the massive spin two is coming from a first derivative of the fundamental field equations, and their linearized version exhibit a novel coupling to the background Weyl tensor.

 This paper is organized as follow. We first introduce the torsion massive gravity model and its field equations (section \ref{sec:model} and \ref{sec3}). We then show how to obtain a non linear version of the FP equation which reduces to the same linear equation in an Einstein background obtained earlier in \cite {Nikiforova:2009qr} (section \ref{sec4}) and examine the coupling to external sources (section \ref{sec5}). Then we turn to the bigravity model (section \ref{sec6}) before concluding (section \ref{sec7}). An appendix gathers some results about the perturbative expansion of the field equations for non propagating degrees of freedom.

\section{Torsion massive gravity}
\label{sec:model}

 We follow the notations of
Refs.~\cite{Hayashi:1979wj,Hayashi:1980ir, Hayashi:1980qp,Nair:2008yh}
and denote the vierbein  by $e_\mu^i$ and the
 connection
%, which can be regarded as an $O(1,3)$ gauge field,
by
 $A_{ij\mu}=-A_{ji\mu}$, where $\mu=0,1,2,3$ and  $i,j=0,1,2,3$
are the space-time and
%index and
% $i,j=0,1,2,3$ are the
tangent space indices, respectively.
We often use the tangent space basis, in which
%$e_\mu^i$ and their inverse $e_i^\mu$.
the indices are
raised and lowered by the Minkowski metric $\eta_{ij}$.
The connection can be
viewed as an $O(1,3)$ gauge field.  It is conveniently
decomposed as follows,
\beqar
A_{ijk} \equiv A_{ij \mu} e^\mu_k &=&  \omega_{ijk} + K_{ijk} ,
\label{con}
\eeqar
where $K_{ijk}$ is called the contorsion tensor and $\omega_{ijk}$ is the spin connection constructed from the vierbein $e_\mu^i$ 
\beq
\omega_{ijk} = \frac{1}{2}\left(C_{ijk} - C_{jik} - C_{kij} \right) \; ,
%\label{T}
\eeq
where
\beq C_{ijk}=e_j^\mu e_k^\nu(\partial_\mu
e_{i\nu}-\partial_\nu e_{i\mu}) = - C_{ikj}.
%\label{C}
\nonumber
\eeq
The spin connection defines a covariant derivative that we denote by $\nabla_l$, while the full covariant derivative using the connection $A_{ij\mu}$ will be denoted by $D_l$. The contorsion tensor is related to the torsion tensor $T_{ijk}=-T_{ikj}$ as follows, 
\beq
K_{ijk}=\frac{1}{2} ( T_{ijk}-T_{jik} - T_{kij})=  - K_{jik}.
\label{K}
\eeq
The torsion tensor can in turn be decomposed into the following  $O(1,3)$ pieces,
% transformations,
%. They are given by
\beq
 T_{ijk}=\frac{2}{3}(t_{ijk}-t_{ikj})+\frac{1}{3}(\eta_{ij}v_k
 -\eta_{ik}v_j)+\varepsilon_{ijkl}a^l,
\label{T'}
%\nonumber
\eeq
where the field $t_{ijk}$ is symmetric with respect to the
interchange of $i$ and $j$ and  satisfies the cyclic and trace
identities,
\beq
t_{ijk}+t_{jki}+t_{kij}=0, \quad\quad\quad
\eta^{ij}t_{ijk}=0,\quad\quad\quad \eta^{ik}t_{ijk}=0.
\label{T''}
\nonumber
\eeq
The 24 independent components of $T_{ijk}$ break up into
4 components of $v_i$, 4 components of $a_i$ and
16 independent components of $t_{ijk}$.

 The model studied in this paper, as well as in Ref.~\cite{Nair:2008yh, Nikiforova:2009qr},
is defined by the action
\beq
S = \int~d^4 x~e~  L ,
\nonumber
\eeq
where  $e= \det e_\mu^i$,
\beq
L=  \frac{3}{2}(\tilde\alpha F - \alpha R)+ c_2 +
c_3F_{ij}F^{ij}+c_4F_{ij}F^{ji}+c_5F^2+c_6(\varepsilon_{ijkl}F^{ijkl})^2,
\label{LF}
\eeq
which uses the curvature $F_{ijmn}$, defined as usual in gauge theories by
\beq
 F_{ijmn}=e_{m}^{\mu}e_{n}^{\nu}(\partial_\mu A_{ij\nu}-\partial_\nu
 A_{ij\mu} +A_{ik\mu}{A^{k}}_{j\nu}-A_{ik\nu}{A^{k}}_{j\mu}),
%\label{curv}
\nonumber
\eeq
the Ricci scalar $R$ constructed from the usual spin connection $\omega_{ijk}$ and 
\beq
F_{jl}=
\eta^{ik}F_{ijkl},\quad\quad\quad F= \eta^{jk}F_{jk},\quad\quad\quad
\varepsilon \cdot F= \varepsilon_{ijkl}F^{ijkl} \; .
%\nonumber
\label{contract}
\eeq

The coefficients appearing in Eq.(\ref{LF}),   $\alpha, \tilde\alpha, c_2, \dots , c_6$ are ``coupling constants'' obeying, apart from
sign restrictions given below,
the only condition
\beq
c_3+c_4+3c_5=0  .
\nonumber
\eeq
In what follows, a combination  $c_1$ of these parameters will be used, defined as 
\beq
c_1 = \frac{3}{2} \left(\tilde{\alpha}- \alpha \right).
\label{kappa-def}
\eeq
For $c_2=0$, the model admits Minkowski space-time
as a solution of the field equations. In that case, the
local $O(1,3)$ invariance is spontaneously broken by the
background value of the vierbein field,
cf. Refs.~\cite{Percacci:1984ai,Percacci:1990wy,Dell:1986pw,Nesti:2007ka,Alexander:2007mt}.
The model with $c_2=0$ is free of ghosts and tachyons in Minkowski background
provided
the parameters  satisfy
inequalities~\cite{Hayashi:1979wj,Hayashi:1980ir,Hayashi:1980qp,  Nikiforova:2009qr},
which in our notations read
\beq
   c_5<0 \; , \quad\quad\quad
c_6>0 \; , \quad\quad\quad
\alpha<0 \; , \quad\quad\quad  \tilde\alpha>0 \; .
\label{cond}
\eeq
Non-vanishing value of $c_2$ enables one to have de~Sitter or
anti-de~Sitter solution with vanishing torsion in this model,
with cosmological constant equal to $\Lambda$. In the latter case,
the requirement of the absence of tachyons imposes one more
 condition~\cite{Nair:2008yh}, i.e.,
\beq
  \tilde{\alpha} > - 4\Lambda c_5 \; .
\label{may6-2}
%\nonumber
\eeq
Since $c_5 < 0$, the latter condition is non-trivial for positive
$\Lambda$. Once the above conditions are satisfied, the theory
is healthy in de~Sitter and
anti-de~Sitter background~\cite{Nair:2008yh}.
Note, for future use, that the curvature Bianchi identity is 
\beq
D_k F_{ijlm} +
{T^n}_{kl}F_{ijmn} + {\rm cyclic}~ (klm)=0.
\label{B1}
\eeq
%where $T_{abc}$ is the torsion tensor.
Contracting this identity twice one obtains
\beq
D^{i} F_{ij} -\frac{1}{2}D_j F
= {T^i}_{kj} {F^{k}}_{i} +\frac{1}{2}{T^i}_{kl}{F^{kl}}_{ji}.
\label{B2}
\eeq

\section{Field equations} \label{sec3}
The dynamical degrees of freedom of the torsion massive gravity model are the vierbein $e^i_\mu$ as well as the torsion $A_{ijk}$ which can in turn be decomposed into $t_{ijk}$, $v_k$ and $a_l$. Here we gather useful results  concerning the field equations of those dynamical degrees of freedom.
\subsection { Gravitational equations}

We obtain the gravitational field equations by setting to zero the variational derivative with respect to 
the vierbein.  The antisymmetric part of these equations is given by
\beqar
\frac{3\tilde\alpha}{2} F_{[ij]} = &= &c_4( F_{[mi]} F_{(mj)}-F_{[mj]}F_{(mi)})
 + \frac{1}{2} (F_{imnj} - F_{jmni}) P_{mn}
 \nonumber\\
  &&- 2c_5F_{[ij]}F
-c_6(\varepsilon_{kmni}F_{kmnj}-  \varepsilon_{kmnj}F_{kmni}) (\varepsilon .F),
\label{g1}
\eeqar
where
\beq
P_{mn}= c_3F_{mn}+c_4F_{nm}.
\label{p}
\eeq
This equation contains no derivative of the curvature tensor  $F_{ijmn}$  and is algebraic. 
It shows that in a flat background the first order $F_{[ij]}$ is zero. In fact it has been shown in  ~\cite{ Nikiforova:2009qr} that this is true in the background of all Einstein manifolds with a vanishing torsion. 
The symmetric part  of the gravitational field equation on the other hand contains explicit derivatives and is given by
\beqar
c_1( F_{(ij)} -\frac{1}{2}\eta_{ij}F) +\frac{\alpha}{\tilde\alpha} (D_k+v_k) H_{k(ij)} 
+\frac{1}{2} ( F_{li}P_{lj} 
-\frac{1}{2}  F_{imnj}+
i\leftrightarrow j )-\frac{1}{2} \eta_{ij} F_{mn} P_{mn}\nonumber\\
+c_5 (2 F_{(ij)} -\frac{1}{2}\eta_{ij} F)F
+c_6 (\varepsilon_{kmni} F_{kmnj} + i\leftrightarrow j -\frac{1}{2}\eta_{ij} \varepsilon .F)(\varepsilon.F)\nonumber\\
+ H_{ij}- \frac{1}{2} \eta_{ij} L_T=0,
\label{g2}
\eeqar
where, 
\beq
H_{ijk}= -\tilde\alpha(t_{kij}-t_{kji})
+\tilde\alpha(\eta_{ki}v_j -\eta_{kj}v_i)
-\frac{3\tilde\alpha}{2}\varepsilon_{ijkl}a^l\;,
\label{Hijk}
\eeq
 and $L_T$ and $H_{ij}$ are given respectively by 
 \beqar
L_T &=& \alpha ( t_{ijk}t^{ijk} -v_i v^i + \frac{9}{4} a^i)\;,
\nonumber \\
\frac{1}{\alpha} H_{ij} &=& 6t_{i(mn)}t_{j(mn)} -\frac{2}{3}t_{i[mn]}t_{j[mn]} -v_iv_j +\frac{3}{4} (\eta_{ij} a^2 - a_i a_j) -\frac{1}{2} (t_{i[mn]} 
  \varepsilon_{jmnp} + i \leftrightarrow j )a_p. \nonumber \\
  \label{1}
\eeqar
Note that only the  term multiplying $c_1$ and the following term involving the derivative contain first order terms in the curvature in the left hand side of Eq. (\ref{g2}). All other terms have second  or higher powers of the curvature.

\subsection{ Torsion field equations}

Setting the variational derivative of S with respect to $A_{ij\mu}$ to zero we obtain the torsion field equations, 
\beqar
-H_{ijk}=\{ \eta_{ik}(D_m P_{jm} -\frac{2}{3} D_j P) - D_i P_{jk} - (i\leftrightarrow j)\}
 +4c_6 \varepsilon_{ijkm} D_m \varepsilon . F + S_{ijk},
\label{torsion}
\eeqar
where, 
\beq
S_{ijk}= \frac{2}{3\tilde\alpha} H_{mnk} (\eta_{im} P_{jn} -\eta_{jm} P_{in} -\frac{2}{3}\eta_{im}\eta_{jn}P + 2c_6 \varepsilon_{ijmn}\varepsilon .F).
\label{s}
\eeq
Equation (\ref{torsion}) can be regarded as a recursive equation for $H_{ijk}$ which can be used to express it in a power series in the components of the curvature tensor.
Note that Equations (\ref{g1}) and (\ref{torsion}) are not completely independent
because of the Bianchi identity.

\subsection{ The $\eta$ and $\varepsilon$ traces of Torsion equation}
Contracting  (\ref{torsion})  with $\eta$  gives, 
\beq
3\tilde\alpha v_i = -2( D_j P_{ij} - \frac{1}{2} D_i P) + S_{ijj}\;,
\label{trH'}
\eeq
where
\beqar
S_{ijj} = \frac{2}{3\tilde\alpha} ( -H_{kij} P_{jk} + H_{njj} P_{in} -\frac{2}{3} H_{ijj}P)\;.
\nonumber
\eeqar
Likewise we can obtain an equation for  $a_l$  by contracting (\ref{torsion}) with $\varepsilon$, 
\beqar
-\frac{3}{2}\tilde\alpha a_l &=&4 c_6 D_l \varepsilon.F - \frac{1}{3} \varepsilon_{ijkl} D_i P_{jk}
\nonumber\\
&&+ \frac{2}{9} ( t_{n[km]} \varepsilon_{kmjl} P_{jn} -v_m \varepsilon_{mljn} P_{jn} - 3 a_jP_{jl} +12 v_l \varepsilon.F)\;.
\label{trtorsion'}
\eeqar

In the Appendix we shall use these equations to show  that in a flat background the trace  and the divergence of $u_{ij}$,  the transverse part of  $a_i$ and the all of $v_i$ will satisfy algebraic equations which can be solved in a perturbative manner order by order in powers of other fields . 

\section{Non linear Fierz-Pauli} \label{sec4}
 We are now in a position to obtain a fully non linear version of Fierz-Pauli theory. This will describe at quadratic order a free massive spin two propagating on the background Einstein space-time, and at non linear order, the full interaction between a massive and a massless graviton as well as a non tachyonic and non ghost-like massive spin zero particle originating from the longitudinal part of the field $a_i$.
Let us first calculate $D_k H_{k(ij)}$  in terms of the curvature components  using the expression of $H_{kij}$ given in  Eq. (\ref{torsion}) and substitute the result in Eq. (\ref{g2}). The result becomes suggestive if we write it in terms of  a new variable $u_{ij}$ defined by
\beq  \nonumber 
u_{ij}= P_{ij} -\frac{1}{6}\eta_{ij} P.
\eeq
We obtain  the following expression \footnote{We recall that  the  covariant derivative  $D$ is with respect to the connection $A_{ij\mu}$ as opposed to  $\nabla$ which is defined relative to the usual spin-connection $\omega_{ij\mu}$ derived from the vierbein $e_{\mu}^i$.},
\beq
 D^2 u_{ij} -  D_i D_m u_{mj} - D_j D_m u_{mi} + \eta_{ij} D_k D_m u_{km} + D_i D_j u - \eta_{ij} D^2 u - m^2 ( u_{ij}-\eta_{ij}u) = \taubis_{ij}
 \label{FP}
 \eeq
where  $m$ is a mass scale,
\beq
m^2= \frac{c_1\tilde\alpha}{3c_5\alpha}
\label{m^2}
\eeq
 and $\taubis_{ij}$  is given by the expression 
\beqar
\taubis_{ij}& = &
 D_m S_{m(ij)} + \eta_{ij} F_{[kl]} P_{[kl]} 
 \nonumber\\
 &&+\frac{1}{2}\{ (F_{ik} -\frac{\tilde\alpha}{\alpha}F_{ki})P_{kj}+ i\leftrightarrow j \} + \frac{1}{2} ( 1+ \frac{\tilde\alpha}{\alpha} ) ( F_{iklj} + i\leftrightarrow j) P_{kl}
 \nonumber\\
 && +\frac{\tilde \alpha}{\alpha}\{ \frac{1}{2}\eta_{ij}(c_2 + F_{mn}P_{mn})+  2c_5 F ( \frac{1}{4} \eta_{ij} F - F_{(ij)})
\nonumber\\
&&  -c_6 (\varepsilon_{mnpi} F_{mnpj} + i\leftrightarrow j - \frac{1}{2}\eta_{ij}\varepsilon.F)\varepsilon.F  \}
\label{tau}
\nonumber\\
&& +\tilde\alpha \{ \frac{1}{2}\eta_{ij} ( t^2 -v^2 +\frac{3}{4} a^2) +\frac{3}{4} a_ia_j -6t_{i(mn)}t_{j(mn)} +\frac{2}{3} t_{i[mn]}t_{j[mn]} 
\nonumber\\
&& 3v_m t_{m(ij)} +\frac{1}{2} (t_{imn}\varepsilon_{jmnp} + i\leftrightarrow j)a_p         \}.
\eeqar
For flat backgrounds and at the linear order  equation (\ref{FP}) reduces to the Fierz-Pauli equation
for a massive spin 2 particle with mass $m$.   More generally, at linear order over an arbitrary Einstein background with vanishing torsion,  the quadratic action which gives rise to equation 
(\ref{FP}) is given by \cite{Nikiforova:2009qr}
\beq
\frac{1}{9c_5^2}S(u,u)= S_{inv}(u,u) + S_m(u,u) + S_W(u,u),
\label{old1}
\eeq
where $S_{inv}$ is the quadratic part of the Einstein-Hilbert action  expanded  around a background Einstein manifold with a Ricci tensor $R_{ij}= 3\Lambda \eta_{ij}$.  It is given by\footnote{Note that for such manifolds the  linear perturbation of $F_{[ij]}$ vanishes. The expression $u_{ij}$ of the present paper is equal to $-3c_5$ times the  corresponding expression $u_{ij}$ of Ref. \cite{Nikiforova:2009qr}.} 
\beqar
S_{inv}( u,u ) =\int d^4 x  \sqrt -\bar g \{-\frac{1}{2} \nabla ^i u^{kl} \nabla_i u_{kl }+ \nabla^i u_{ki} \nabla_l u^{lk} -  \nabla ^l u \nabla^k u_{lk} 
\nonumber\\
+\frac{1}{2} \nabla ^i u \nabla_i u  - W_{iklj} u^{ij} u^{kl}  - \Lambda (u^{kl} u_{kl} + \frac{1}{2} u^2) \}\;.
\label{biE}
\eeqar
The two other terms  in (\ref{old1}) are defined by
\beq
S_m(u,u)= -\frac{\tilde M^2}{2}\int \sqrt -  g ( u_{ij}u^{ij} - u^2)\;,
\label{quad}
\eeq
and
\beq
 S_W(u,u)= \frac{s}{2} \int \sqrt - g  W_{iklj} u^{ij} u^{kl},
\label{old2}
\eeq
where $\tilde M^2$ and $s$ are parameters defined in  \cite{Nikiforova:2009qr}  and $W_{iklj}$ is the Weyl tensor of the background manifold.  The latter tensor has all the symmetries of Riemann tensor and is trace free on all pair of indices.  For an Einstein metric a Bianchi identity implies that the $\nabla^i W_{ijkl}=0$. 

We shall see later that for a class of  bigravity models  a linear combination of the perturbations for the two metrics over an Einstein space-time background has an action analogous to equations (\ref{old1})-(\ref{old2}) with $s=0$. It should, however,  be noted that our generalized non-linear Fierz-Pauli equation  (\ref{FP})  is written for a linear combination of the curvature tensor rather than the metric. Thus the field equations in terms of the fundamental fields will be higher than second order. Nevertheless a detailed analysis in the presence of the external sources indicates that the poles in the propagator are only second order simple poles with non tachyonic residues. We shall review this briefly in the following.

\section{Coupling to external sources} \label{sec5}

One of the  interesting features of the torsion gravity model, with possible observational consequences, emerges if we couple the vierbein fluctuations around a flat background to conserved external  symmetric energy momentum tensor.  It is usually assumed that in order to have non zero torsion one needs to have spinning matter in the universe.  The previous analysis of our models shows that this is not the case\footnote{ The general case of coupling to external torsion sources as well as vierbein sources have been given in our notation in \cite{Nikiforova:2009qr}.}. In fact in these models, even in the absence of spinning sources, one can excite torsion degrees of freedom by coupling to conserved symmetric second rank energy momentum tensor $\tau_{ij}$.  Moreover, the fact that the nonlinear Fierz-Pauli equation (\ref{FP}) is in fact a third order differential equation for the components of the connection does not lead to any pathologies in the propagator, as we will see. It has been shown in a previous paper \cite{Nikiforova:2009qr} that the vierbein excitations sourced by $\tau_{ij}$ are given by
\beq
h_{ij}=
\frac{1}{c_1} \; \frac{1}{k^2} \left( \tau_{ij} -\frac{1}{2}\eta_{ij} \tau
\right) - \frac{\tilde\alpha}{\alpha c_1} \frac{
1}{k^2+m^2}\left( \tau_{ij}-\frac{1}{3}\eta_{ij} \tau \right),
\label{f19}
\eeq
where,   $m$ is the mass of the spin-2 particle  given by Eq. (\ref{m^2}). 
The presence of the massive spin-2 pole is a reflection of the non vanishing torsion.  In fact the fluctuations of the connection components are given by, 
 \begin{align*}
A_{ijk} = & - \frac{i}{c_1} \frac{1}{k^2} \left\{
k_i \left( \tau_{jk} -\frac{1}{2}\eta_{jk} \tau
\right) - k_j \left( \tau_{ik} -\frac{1}{2}\eta_{ik} \tau
\right) \right\}
\\
& + \frac{i}{c_1} \frac{
1}{k^2+m^2}
 \left\{
k_i \left( \tau_{jk} -\frac{1}{3}\eta_{jk} \tau
\right) - k_j \left( \tau_{ik} -\frac{1}{3}\eta_{ik} \tau
\right) \right\}.
%\nonumber
\end{align*}
These components are different from the Riemannian connection corresponding to
the vierbein perturbation (\ref{f19}) which clearly shows mixing between vierbein and
torsion fields in our model.

We see from  (\ref{f19}) that the coupling of the massless and massive spin-2 particles to the  energy-momentum tensor are independent from each other. This can lead to some interesting phenomenology. In particular it allows to adjust the parameters  of the torsion massive gravity in order to have both the Non Relativistic-Non Relativistic and Non-Relativitic-Relativistic interactions to agree with standard predictions of GR. This is just because we now have two gravitons at hands and contrasts with the theory of a single massive graviton where, if Cavendish experiments agrees with the predictions of GR, light bending would differ.

 \section {Bigravity} \label{sec6}
 Some versions of bigravity are known to have a spectrum on flat space-time, containing one massless and one massive graviton. Here we underline that the same is true for Einstein space-times which can always be made into a background solution of the kind of bigravity we consider, at the price of some tuning in the action. To see this, consider the action {\bf 
 \beqar \label{ACTBI}
 S= \frac{M_g^2}{2}\int d^4 x \sqrt{-g} ( R(g) -2 \Lambda_g) + \frac{M_f^2}{2}\int d^4 x \sqrt{-f} ( R(f) -2 \Lambda_f) 
 \nonumber\\
 -\frac{m^2M_g^2}{8}\int d^4 x \sqrt{-g} V({\bf g^{-1} f}) \;,
 \label{bi}
 \eeqar
 } where $g_{\mu\nu}$ and $f_{\mu\nu}$ are two independent metrics on the same manifold  and $ V$ is some scalar function built out of invariants made from the "matrix" ${\mathcal M}^{\mu}_{\hphantom{\mu} \rho} \equiv g^{\mu \nu} f_{\nu \rho}$. Such theories were first considered in the context of strong interactions \cite{Isham:gm,Salam:1976as,Isham:1977rj}, and more recently were revisited in a cosmological context \cite{Damour:2002ws,Damour:2002gp,Damour:2002wu,Blas:2007zz,Blas:2005yk,Berezhiani:2007zf,Blas:2009my}. The action is clearly invariant under general coordinate transformations and it
 is not hard to see that the equations of motion have a solution such that  $g_{\mu\nu}=f_{\mu\nu}$ and 
\beqar
R_{\mu\nu} (g) = R_{\mu\nu} (f) =  3\Lambda g_{\mu\nu},
\eeqar
where the cosmological constant $\Lambda$ of the Einstein manifold is related to the parameters entering the action through  the relations
\beqar
 \Lambda_g &=&  3\Lambda -\frac{1}{8} m^2 \bar V + \frac{1}{4}m^2 v,
 \label{1} \\
 \Lambda_f &=& 3\Lambda -  \frac{1}{4}m^2 v \frac{M_g^2}{M^2_f}.
 \label{2}
\eeqar
Here  $\bar V$ and $v$ are  defined by  expanding V around the background solution obeying $\bar g_{\mu\nu}= \bar f_{\mu\nu}$, and hence also ${\cal M}^{\mu}_\nu = \delta^\mu_\nu$. One has 
\beqar
V= \bar V+ v\delta^\mu_\nu \delta  {{\mathcal M}^{\nu}}_{\mu} +\frac{1}{2}( w_1 \delta^\mu_\nu \delta^\lambda_\sigma + w_2 \delta^\mu_\sigma \delta^\lambda_\nu)  \delta  {{\mathcal M}^{\nu}}_{\mu}  \delta  {{\mathcal M}^{\sigma}}_{\lambda} + \cdots  
 \eeqar
 where $\delta \mathcal M^{\mu}_\nu$ refers to the variation of $\mathcal M^{\mu}_\nu$ and we stopped the above expansion  at second order.
We can expand the action up to bilinear terms to find the spectrum around  the background solution.  To this end we write 
\beqar
g_{\mu\nu} &=& \bar g_{\mu\nu} + \alpha h_{\mu\nu} + \beta u_{\mu\nu}, \\
f_{\mu\nu} &=& \bar g_{\mu\nu} + \alpha' h_{\mu\nu}+ \beta' u_{\mu\nu},
\eeqar
 where $\bar g_{\mu\nu}$ denotes the background Einstein metric and $h_{\mu\nu}$ and $u_{\mu\nu}$ are metric fluctuations.  Expanding  the action up to quadratic terms 
 we find out that it becomes diagonal in terms of the fields $h_{\mu\nu}$ and $u_{\mu\nu}$, provided one chooses $\alpha=\alpha'$ and $\beta' = - \frac{M_g^2}{M_f^2}\beta$.  The field $u_{\mu \nu}$ is then seen to represent a massive graviton which will have a standard Fierz-Pauli mass term { provided we constrain  the potential $V$ to obey $v= -2 \left(w_2+w_1\right)$. Upon a rescaling of $u$
 of the form  $ u\rightarrow  2 u {(\beta^2 M_g^2 + {\beta' }^2 M_f^2)}^{-{\frac{1}{2}}}$  the bilinear action then reduces to 
\beq
S_{bigravity}= \frac{M_{pl}^2 }{2}S_{inv}(h,h)+S_{inv}(u,u)+ S_m(u,u),
\label{new1}
\eeq
where  $ S_{inv}$ is the bilinear part of the Einstein -Hilbert action given explicitly in (\ref{biE}),
\beq
M_{pl}^2 = \alpha^2 \frac {M_g^2 + M_f^2}{2} \;,
\label{pl}
\eeq
 and $S_{m}(u,u)$ is given by (\ref{quad}) where $\tilde M^2$ is  defined by
\beq
\tilde M^2 =  \frac {M_g^2 + M_f^2}{2M_f^2}m^2 (\frac{v}{2} -w_1)\;.
\label{pl}
\eeq
As in manifest in Eq.(\ref{new1}) the field $h_{\mu\nu}$ is describing a massless graviton (whose presence can be attributed to the invariance of action (\ref{ACTBI}) under diffeomorphisms).
 We note that the action given by Eq.(\ref{new1}) can formally be obtained from the one of the torsion model (\ref{LF}) by setting $s=0$ in eq. (\ref{old1}).
 
The field equations for the massive graviton $u$ field are easily obtained and read 
\begin{align}
 \nabla^2 u_{\mu\nu} -\nabla_\mu\nabla^\lambda u_{\lambda\nu}
-&\nabla_\nu\nabla^\lambda u_{\lambda\mu}+\nabla_\mu\nabla_\nu u
+\bar g_{\mu\nu}\left(\nabla^\kappa\nabla^\lambda u_{\kappa\lambda}-\nabla^2 u\right)
-2\Lambda\left(u_{\mu\nu}+\frac{1}{2}\bar g_{\mu\nu}u\right)
\nonumber \\
&
-\tilde M^2(u_{\mu\nu}-\bar g_{\mu\nu}u)
-2W_{\mu\kappa\lambda\nu}u^{\kappa\lambda}=0\;,
\label{FP'}
\end{align}
where all the indices are raised and lowered by the background metric $\bar g_{\mu\nu}$, the covariant derivatives are computed using the connection derived from $\bar g_{\mu\nu}$ and $ W_{\lambda\mu\nu\sigma}$ is the Weyl tensor of the background metric. The divergence of this equation yields, 
\beq
\tilde M^2 \nabla^\mu( u_{\mu\nu} - \bar g_{\mu\nu} u)=0\;.
\label{div|}
\eeq
We make use this equation in  the trace of (\ref{FP'}) to obtain
\beq
-3(2\Lambda +\tilde M^2)u=0
\label{tr}
\eeq
From the last two equations we thus see that if  $2\Lambda +\tilde M^2$ is different from zero  both $u$ and $\nabla^\mu u_{\mu\nu}$ will vanish. In this case (\ref {FP'}) reduces to 
  \begin{align}
 \nabla^2 u^{tt}_{\mu\nu} 
-(2\Lambda +\tilde M^2 )u^{tt}_{\mu\nu}
-2{{W_{\mu}}^{\kappa\lambda}}_\nu u^{tt}_{\kappa\lambda}=0
\label{FP''}
\end{align}
 where $u^{tt}_{ij}$ indicate transverse traceless part of $u_{ij}$. We thus see that we have only 5 propagating degrees of freedom,  the same as in linear FP equation.

 \section{Conclusion and Outlook} \label{sec7}
 In this work, we have introduced a fully non linear version of the Fierz-Pauli theory for a massive spin two, based on a theory with propagating torsion. This "torsion massive gravity" has several interesting features.  At the quadratic level and around a flat space or a weak field Einstein manifold, it describes a healthy spectrum consisting of a massive spin-zero particle, a massless, and a massive graviton. At higher order, one can extract from this model a fully non linear version of the linear Fierz-Pauli field equation. This non linear equation describes the propagation and interactions of the massive spin two on any curved background space-time, provided the latter is a solution of the theory. We also showed that a similar spectrum  of spin-2 particles arises in some chosen bigravities and in the same kind of backgrounds.  However, there are also important differences between those bigravities and the torsion massive gravity. First, an intriguing feature is that the generalized Fierz-Pauli field equation is satisfied by a linear combination of the components of the curvature tensor. It may therefore be worrisome  that, being a third order differential equation in terms of the connection components, it can give rise to ghosts and tachyons. However, this does not happen.  It does not generate higher than second order poles when one couples the fundamental fields, the vierbein and the connection, to appropriate external sources. This is in contrast with the bigravity theory where the massive spin two equation is second order to begin with. An issue which we did not address and leave for future work, is the investigation of the  Boulware-Deser ghost in a non perturbative framework. In massive gravity with finite number of degrees of freedom, it is indeed expected that a ghost, or an extra degree of freedom, if not present at linear order, will start to propagate at non linear level. The Hamiltonian counting of degrees of freedom for bigravity \cite{Damour:2002ws} suggests that it indeed happens in these models\footnote{Thus the non perturbative spectrum of bigravity coincides with that of torsion model at the linearized order, both models propagate a total of 8 physical degrees of freedom}. The pertubative expansion done in this paper does not allow to settle this issue,  even though the calculations given in the appendix, showing that fields which are non propagating at linear order can be expressed alebraically in terms of propagating fields up to quadratic order, can be seen as encouraging. Note, however, that the fact that the massive graviton in our model is a derivative of fundamental fields might be of crucial importance for this. Another interesting difference between bigravity and torsion massive gravity is that our generalized Fierz-Pauli equation in an Einstein background contains a term not present in bigravity, involving the background Weyl tensor.  Finally, since the bigravity model is obtained from a manifestly coordinate invariant action, the inclusion of higher order terms in a perturbative expansion will  also produce a non linear Fierz-Pauli equation incorporating the interactions of the massive and massless gravitons. At the linear level on an  arbitrary Einstein background the two models produce two FP-type equations. Clearly the non linear versions derived from them are also expected to be different.

\section*{Acknowledgements}
S.R.D acknowledges hospitality at the APC laboratory  where this collaboration started. 
C.D. thanks for its hospitality the Abdus Salam International Centre
for Theoretical Physics, Trieste, where part of this work was done. 

%\newpage
%\section*{Appendix A: Background manifolds}
%\def\theequation{A\arabic{equation}} \setcounter{equation}{0}

\begin{appendix}
\section{Expansion of non propagating fields}
In this Appendix we show that in a flat background some fields can be eliminated in terms other fields in an order by order manner in a perturbative expansion in powers of the fields.

Firstly by inspection of Eq. (\ref{g1}) it is obvious that $F_{[ij]}$ is non zero only at the quadratic order or higher. Therefore, up to the second order, terms of the form $F_{[ij]}$ times any other field will be third order or higher. Such terms can be neglected if we are studying equations at most up to the second order.  In other words we can write Eq. (\ref{g1}) as 
\beqar
\frac{3\tilde\alpha}{2} F_{[ij]} = 
 + \frac{1}{2} (F_{imnj} - F_{jmni}) P_{(mn)}
 -c_6(\varepsilon_{kmni}F_{kmnj}-  \varepsilon_{kmnj}F_{kmni}) (\varepsilon .F) + ...
\label{g1'}
\eeqar
where above $...$ refer to the 3rd order terms or higher.  

Next consider Eq. (\ref{trH'}).  In this equation  the contribution of  $S_{ijj}$ to  $v_i$  is manifestly of the form of the product of two fields.  We want to argue that the same is true also for the terms inside the parenthesis on the right and side of (\ref{trH'}).  To see this, first we decompose $ P_{ij}= P_{(ij)} + P_{[ij]}$, where  $P_{[ij]} = (c_3-c_4) F_{[ij]}$ and  and $P_{(ij)}= -3c_5 F_{(ij)}$.  We have already argued that $F_{[ij]}$ starts from the second order terms. Thus we need only to concentrate on the contribution of $ D_j P_{(ij)}- \frac {1}{2} D_i P = -3c_5 ( D_j F_{(ij)}- \frac {1}{2} D_i F)$ part.  For this it is useful to use the Bianchi identity (\ref{B2}) in which the right hand side is clearly quadratic in the fields. Decomposing $F_{ij}= F_{(ij)} + F_{[ij]}$  and remembering again that $F_{[ij]}$ is bilinear in the fields we conclude that so is also 
 $( D_j F_{(ij)}- \frac {1}{2} D_i F)$. This completes the proof that the right hand side of (\ref{trH'}) is quadratic in the fields.  Thus $v_i$ is second order in the fields which agrees with the fact that it vanishes at the linear order on flat backgrounds.  This result is important because it shows that any term like $v_i $ times any other field on the right hand side of Eq. (\ref{trH'}) will be third order or higher and thus can be neglected up to the order of interest to us. Since there are no linear terms in $v_i$ on the right hand side of  eq. (\ref{trH'}) we can regard this equation as an algebraic equation determining $v_i$ as a second order polynomial of other fields.

 Next  let us consider $u =  \frac{1}{3} P = - c_5 F$.  There are several expressions for this object, but the most convenient is the  exact  relation which can be obtained from taking the trace of Eq. (\ref{g2}).  A simple calculation shows that  one has 
\beq
  u= \frac{c_5\alpha}{c_1} ( t^2 - v^2 +\frac{9}{4} a^2 + 3 \nabla.v).
 \label{F}
 \eeq
Since we have already established that $v$ is quadratic in the  fields, the right hand side above obviously starts from  quadratic terms. The important point to note is that $\nabla.v$ does not contain $u$ itself. 
This can be seen from investigating the expression for $v_i$ given in (\ref{trH'}).  Thus upon substituting for $v_i$ on the right hand side of (\ref{trH'}) we shall obtain an expression which expresses $u$ algebraically in terms of the remaining fields. 

Thus far we have shown that up to the second order terms we can express $u$ and $v_i$ algebraically in terms of other fields.   We can use these facts in the Bianchi identity (\ref {B2}) to deduce that $D_i F_{(ij)}$ is second order in the fields and is given algebraically as the product of other fields not involving $u$ or $v_i$ or $\partial _i u_{ij}$. 

Finally we need to examine the propagation of  $a_i$.  We know that at the linear level it is only the longitudinal part of $a_i$ which propagates.  For a detailed study of the dynamics of this field we give an expanded form for $\varepsilon.F$, viz, 
\beq
\varepsilon.F = 6 \nabla.a + \frac{8}{9} \varepsilon_{ikjl}t_{m[ik]}t_{m[jl]} - 4 a.v\;.
\label{e.F}
\eeq
This is an exact expression.  We can drop the $a.v$ term as it is third order or higher.  In the first term we can expand in powers of the metric perturbations $h^{i\mu}$ defined by 

\beq
e^{i\mu}= \eta^{i\mu} + h^{i\mu}.
\label{vier}
\eeq
 We thus obtain, 
 \beq
\varepsilon.F =6 \partial.a +6 h^{ij} \partial_i a_j - \omega^L_{jii} a^j+.....+ \frac{8}{9} \varepsilon_{ikjl}t_{m[ik]}t_{m[jl]},
^L_{jii} a^j  \label{e.F'}
 \eeq
 where $\omega^L_{ijk} = -\omega^L_{jik}$ is the linear part of the spin connection $\omega_{ijk}$ which can be expressed in terms of $h^{ij}$ and $...$ denote third  or higher powers of the fields. We see that apart from the first term on the right hand side all other terms have two or more powers of the fields.  Let us now consider Eq.(\ref{trtorsion'}).  First we note that the first term on the right hand side of this equation (in which $D_l= \partial_l$) is always a longitudinal object.  In the second term in this equation it is the antisymmetric piece $P_{[ij]}= (c_3-c_4)F_{[ij]}$ which contributes, which itself is quadratic in the fields. The remaining terms are manifestly quadratic or higher in the fields. If we reiterate this equation we see that up to the second order terms the right hand side will involve only the longitudinal piece, the transverse part being second order will not contribute in the product of two fields. The transverse part of $a_i$ is given by
 \beq
 a^t_l = \frac{2}{9}\frac{c_3-c_4}{\tilde \alpha} \varepsilon _{ijkl} D_i F_{[jk]} + \frac{4}{9}\frac{c_5}{\tilde \alpha} ( \varepsilon _{kmjl} t_{n[km]} f_{(jn)} - 3 a_j F_{(jl)})^t +...
 \label{at}
 \eeq
 where the superscript  $t$ indicates transverse part.  This shows explicitly that $a_i^t$ is algebraic and is given as a bilinear in other fields. On the right hand side of the above equation we should, of course replace $a_j$ by its longitudinal part. 
 \end{appendix}

\end{document}